\documentclass[conference,compsoc]{IEEEtran}
\usepackage[compress]{cite}
\usepackage{float}
\usepackage{graphicx}
\usepackage{latexsym}
\usepackage{amssymb}
\usepackage[fleqn]{amsmath}
\usepackage{color}
\usepackage[letterpaper,left=1in,top=1in,bottom=1in,right=1in,includeheadfoot,headheight=.6pt]{geometry}

\IEEEoverridecommandlockouts

\begin{document}
\title{Quantum Accelerators for High-Performance Computing Systems}

\author{
\IEEEauthorblockN{Keith A.~Britt, Fahd A.~Mohiyaddin, and Travis S.~Humble\IEEEauthorrefmark{2}} 
\IEEEauthorblockA{Quantum Computing Institute\\
Oak Ridge National Laboratory\\
Oak Ridge, Tennessee USA}
Email: \IEEEauthorrefmark{2}humblets@ornl.gov
\thanks{This manuscript has been authored by UT-Battelle, LLC, under Contract No. DE-AC0500OR22725 with the U.S. Department of Energy. The United States Government retains and the publisher, by accepting the article for publication, acknowledges that the United States Government retains a non-exclusive, paid-up, irrevocable, world-wide license to publish or reproduce the published form of this manuscript, or allow others to do so, for the United States Government purposes. The Department of Energy will provide public access to these results of federally sponsored research in accordance with the DOE Public Access Plan (http://energy.gov/downloads/doe-public-access-plan).}
}

\maketitle 

\begin{abstract}
We define some of the programming and system-level challenges facing the application of quantum processing to high-performance computing. Alongside barriers to physical integration, prominent differences in the execution of quantum and conventional programs challenges the intersection of these computational models. Following a brief overview of the state of the art, we discuss recent advances in programming and execution models for hybrid quantum-classical computing. We discuss a novel quantum-accelerator framework that uses specialized kernels to offload select workloads while integrating with existing computing infrastructure. We elaborate on the role of the host operating system to manage these unique accelerator resources, the prospects for deploying quantum modules, and the requirements placed on the language hierarchy connecting these different system components. We draw on recent advances in the modeling and simulation of quantum computing systems with the development of architectures for hybrid high-performance computing systems and the realization of software stacks for controlling quantum devices. Finally, we present simulation results that describe the expected system-level behavior of high-performance computing systems composed from compute nodes with quantum processing units. We describe performance for these hybrid systems in terms of time-to-solution, accuracy, and energy consumption, and we use simple application examples to estimate the performance advantage of quantum acceleration.
\end{abstract}

\section{Introduction}
Many advantages of quantum computing are now well understood at the algorithmic level, but whether quantum computing can support practical applications remains an open question \cite{mohseni2017commercialize}. For example, despite enormous potential to support scientific discovery \cite{ascrreport}, there are many technical challenges to the integration of quantum algorithms into practical HPC system design \cite{britt2017high}. Foremost is the realization of quantum processing units (QPUs) that are sufficiently large in capacity and robust in operation to provide reliable solutions for problem of practical significance. In this regard, current QPUs are little more than toys that enable exploratory research and designs, though there is rapid progress toward QPUs capable of so-called quantum supremacy with some estimates for this crossover placed as early as 2018 \cite{Lund2017}.
\par
Given the apparent slowdown in Moore's law and the implication that future performance gains must come from alternative technologies \cite{Conte2015}, the potential for QPUs to impact existing applications has never been more significant. As a specific and practical use case, we explore the relevance of QPUs to accelerate high-performance computing (HPC) and especially in the role as computational accelerators \cite{humble2016software,Svore2016}. We define some of the programming and system-level challenges facing the integration of QPUs to HPC, and we discuss a novel quantum-accelerator framework that uses specialized kernels to offload select workloads while also maintain existing computing infrastructure.
\section{State of the Art}
State-of-the-art scientific computing, especially for large-scale applications, lies in the massively parallel heterogeneous architectures expressed by modern HPC systems. The TOP500, for example, highlights machines that quickly solve a suite of benchmark problems \cite{Top500}. The dominate strategy to winning this competition relies on parallelism at the node, thread, and instruction level parallelism with various hardware and software strategies for realizing each approach. Current systems are routinely benchmarked over $10^{18}$ floating-point operations per second, or 1 petaflop. Future designs for exaflop systems are now being made with expected delivery dates of 2021, if not sooner. 
\par
As part of the effort to maximize the parallelism of HPC systems for scientific computing, attention has been placed on node designs that incorporate specialized co-processors acting as accelerators \cite{kurzak2010scientific}. The purpose of the accelerator is to solve select computational tasks more efficiently using a combination of device design and programming optimization. Currently, one of the most prominent examples of an accelerator is the GPU \cite{mittal2015survey}, which is an integral part of the US Titan supercomputer at Oak Ridge National Laboratory. GPUs offer efficient implementations of SIMD programs (at a per thread level) that map particularly well onto common scientific calculations, e.g., matrix-vector operations. However, effective usage of the accelerator paradigm requires the judicious selection of which tasks should be off-loaded and to which accelerator device.
\par
We consider how the HPC accelerator architecture can be adopted to include quantum processing units (QPUs). Our motivation is twofold. First, trends in the design of HPC architectures suggest that the accelerator paradigm is likely to persist into the exascale period. Future HPC designs are anticipated to focus on extreme-scale heterogeneity, in which multiple accelerators may be allocated within a node. The impact of QPUs on scientific computing is therefore greater if they can be reconciled with these prevailing machine designs. Second, existing application stacks for scientific computing are complex hierarchies of concerns and abstractions. The holistic rewriting of such software is not only ineffective but also unlikely. In order to maximize the benefit of QPUs to the scientific user, it is therefore advantageous to integrate these processors into existing applications through abstractions of specialized functionality. Similar to how GPUs accelerate low-level linear algebra methods, we expect QPUs may offer boost to applied mathematical tasks. Of course, QPUs currently represent a very risky and technically challenging technology and it is likely to require several years to develop a mature HPC infrastructure that support these ideas.
\par
\begin{figure}
\centering
\includegraphics[width=\columnwidth]{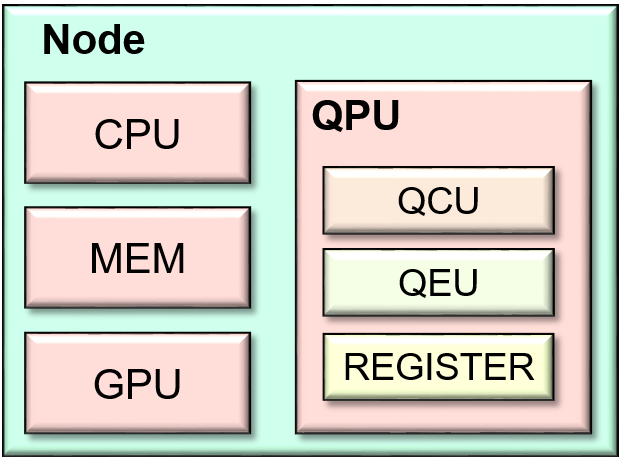}
\caption{A typical node within the QPU-accelerated HPC system}
\label{fig:node}
\end{figure}
\par
Given the remarkable prowess of current HPC systems, it may be difficult to appreciate the relevance of QPUs to scientific computing as accelerators \cite{feynman1982simulating}. For more than 25 years, the global scientific community has pushed the development of quantum technology toward more relevant scales. The fundamental principles of quantum computing, namely, superposition, entanglement, and randomness, have long been demonstrated in proof-of-principle experiments. But the translation of these physical concepts into a meaningful technology has been a tremendous challenge due to the extreme sensitivity of these physical systems to underlying noise \cite{jones2012layered}. State-of-the-art chips from IBM and Google have 16 to 22 qubits and near-term plans of 50-qubits may seem like meager improvements. But the computational power of a QPU does not scale linearly with its capacity, i.e., number of qubits. Rather the growth is exponential due to the principle of quantum superposition. A jump of more than 20 qubits in capacity therefore equates with a $2^{20}$ jump in computational power. Therefore applications with 50-qubit QPUs may be able to outperform their conventional counterparts. Tests of quantum supremacy at these scales are limited to seemingly contrived problems, for example, generating chaotic sequences of truly random numbers. Future tests must be applied to more practical problem sets in scientific computing including computational chemistry, materials design, and machine learning.
\par
Despite the success made in developing QPUs, there are multiple barriers to the integration expected between these devices and conventional HPC systems. Technological barriers arise from several factors. First is the consequence that most quantum computing technologies are not based on CMOS technology. There are several approaches toward silicon-based quantum computing, but even those are not directly compatible with existing processors. Second, nearly every quantum technology requires some level of thermodynamic control to manage errors. For quantum computation, this control manifests as isolation by ultra-high vacuum and cryogenic refrigeration systems. These sensitive control systems are largely incompatible with the noise and vibrations found in a modern server room.
\par
A less obvious, but equally significant, barrier to integration is the logical interactions between the conventional and quantum processing systems. Conventional processors are managed and accessed by dedicated operating systems, which have often been highly specialized to standardized execution and programming models. By contrast, QPUs are currently managed by event-driven programs that interface with one-of-kind control systems consisting of field generators. Designing software that accounts for the tremendous differences between these hardware systems is an ongoing challenge that changes continuously with improvements in device physics and use cases.
\section{Quantum Accelerator Design}
We describe a system-level model for the integration of QPU's as accelerators into HPC systems \cite{britt2017high}. Our approach is based on the use of these quantum accelerators for supporting large-scale scientific computations. We elaborate on the role of the host operating system to manage these unique accelerator resources, the prospects for deploying quantum modules, and the requirements placed on the language hierarchy connecting these different system components. 
\par
\begin{figure}
\centering
\includegraphics[width=0.8\columnwidth]{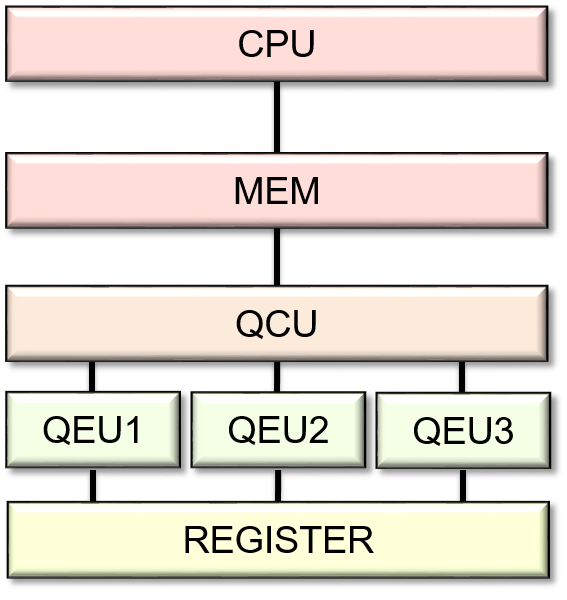}
\caption{A component diagram defining the interfaces within a QPU-accelerated compute node.}
\label{fig:stru}
\end{figure}

We first describe the structure of the QPU in terms of three different subcomponents. As shown in Fig.~\ref{fig:node}, the QPU consists of a quantum control unit (QCU), quantum execution units (QEUs), and a quantum register. The QCU represents the interface between the QPU and the external system. The role of the QCU is to parse incoming instructions and issue operands to the various QEUs. Each QEU represents an implementation of a subset of available gate operations. Multiple QEUs may be needed to express all possible gate operations and these may have the benefit of operating in parallel. Figure \ref{fig:stru} emphasizes that each QEU acts on the quantum register by emitting an applied field. The field itself will depend strongly on the underlying technology, the intended gate operation, and any error mitigation methods. Notably, the QEU represents the conversion of the digital operand into an analog field, and the interaction between the field and the register is modeled using quantum mechanical formalism. The simplest such model would describe the interaction using a Hamiltonian form. 
\par
\par
As a physical subcomponent, the quantum register represents where quantum information is stored within the QPU. The register itself is composed of quantum register elements - in isolation, each register element stores a single qubit of information. Of course, $n$ register elements may store $n$ separable qubits of information as well as an infinite number of non-separable states. For simplicity, we refer to an $n$-qubit register as having a capacity of $n$ qubits. The interactions between the register and the QEUs work in both direction. Actions on the register may also generate a field following the application of a gate field by a QEU. The most obvious example is during measurement readout, which will cause the register to output a field indicating the collapse of some register elements to a state in the measurement basis. The resulting field must be collected by the appropriate QEU and then converted to a digital signal. Following analog-to-digital conversion, the information is relayed to the QCU, which may either forward the results back to the host system, or collect the information for further processing, e.g., syndrome decoding as part of fault-tolerant protocols.
\section{Language Hierarchy}
\begin{figure}
\centering
\includegraphics[width=0.9\columnwidth]{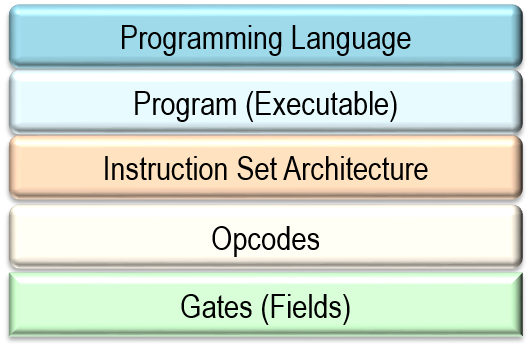}
\caption{The language hierarchy used to program and control a QPU.}
\label{fig:lang}
\end{figure}

The structure described above imposes a natural language hierarchy on the operation of the QPU. This hierarchy is summarized in Fig.~\ref{fig:lang}. We have described the QPU component in terms of its subcomponents and its interfaces. This component is understood to be part of a larger system structure, the host HPC system. The interactions between the host and accelerator devices are driven by user-defined programming statements. We discuss some of the languages required to manage the interactions between a conventional host program and a QPU device.
\par
\par
As shown in Fig.~\ref{fig:lang}, the highest level language is a typical programming language. However, in order for a user to access the unique capabilities offered by a QPU, the programming language must expose some artifacts, such as keywords, data types, or library functions, that permit quantum logical operations. Depending on the experience of the user, they may either use existing programming languages augmented with specialized libraries (with appropriate bindings) or dedicated quantum programming languages, of which several now exists \cite{Selinger2004,Abhari2012,Green2013,Wecker2014,steiger2016projectq}. However, the purpose of the prepared source code is to undergo transformation into an executable that can access and control of the accelerator device, along with other system devices and resources. Currently, there are no such executable programs for QPUs. Rather every existing QPU prototype is managed through an interpreted environment. While these environments, such as Python and Lisp, are capable of running scripts, they do not address the more complicated question that arise within a dedicated management system. We must expect that more mature realization of QPUs will require a specialized language for expressing quantum executables that can communicated with the host OS.
\par
Whether the executable is interpreted or not, the role of the host is to issue instructions to the device \cite{smith2016practical,cross2017open}. The instruction set architecture (ISA) is an important element in the design of the QPU as it impacts both the compiled representation of the program as well as the efficiency of the underlying device operations \cite{britt2017instruction}. The ISA is an abstraction of the device capabilities that must include appropriate constraints on possible operations. For example, most QPUs have significant limits on connectivity, i.e., constraints on which qubits can undergo simultaneous operations. Connectivity constraints can often require additional movement or teleportation operations to co-locate data in appropriate register location. We expect that the host compiler will ultimately have responsible for satisfying these constraints during executable construction, but the ISA must provide sufficient information for this purpose. Similar, different quantum technologies support unique operations and `cross-compiling' for different QPU will require specific knowledge about the support ISA.
\par
The stream of instructions and data sent to the QPU must next be parsed into the opcodes that will trigger specific quantum execution units (QEUs). Again, technology-specific limitations as well as performance tunings are likely to influence the design of these opcodes. For example, a long-term goal of quantum computing is to sustain fault-tolerant computations for arbitrarily long programs. This requires the use of quantum error correction and fault-tolerant (FTQEC) operations, which may manifest in device specific implementations. The FTQEC opcodes of a specific QPU may be tuned to ensure negligible errors rates.
\par
Finally, the specification for how QEUs implement the opcodes by issuing fields to the available register represents a level of language that is unlikely to be seen by the user. However, it is necessary to provide specifications of this layer in order to realize the preceding layers as well as to provide methods for device designers to evaluate instruction side effects and counter on-chip noise.
\par
\begin{figure}
\includegraphics[width=\columnwidth]{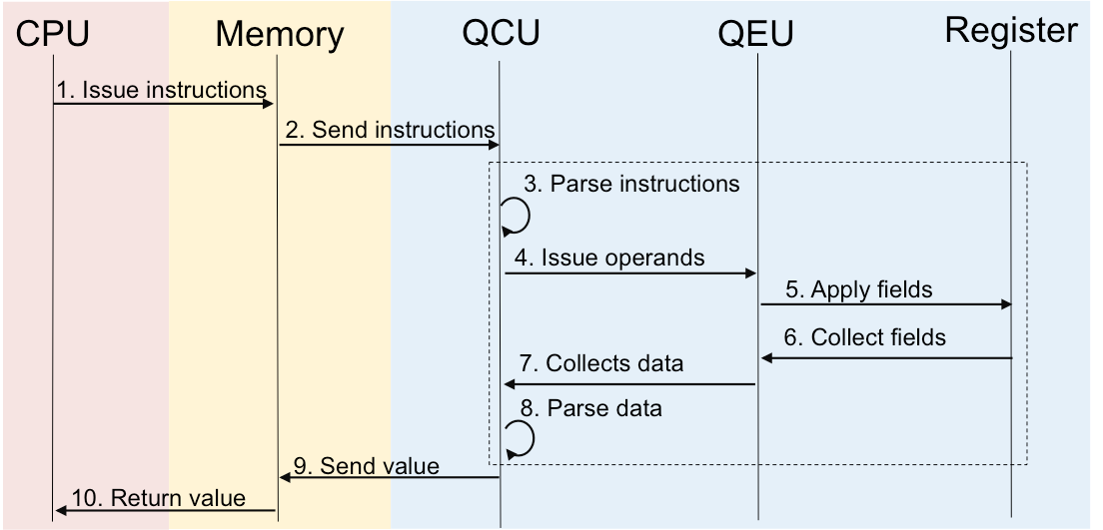}
\caption{The execution model for a QPU-accelerated host node passing instruction from a CPU to memory to a QPU.}
\label{fig:exe}
\end{figure}

We provide a simple example of the execution model needed to realize this language hierarchy in Fig.~\ref{fig:exe}. This model emphasizes the interaction between the host and device as well as the potential for the QPU to loop through repeated instructions before responding. We have assumed that these interactions are mediated via the system memory and appropriate memory controllers, but other possibilities include the current client-server paradigm that is used by every QPU prototype presently.
\section{System Model and Simulation}
As an example of how a QPU may be used within an accelerator architecture, we have made a preliminary system model using the Structural Simulation Toolkit (SST) \cite{rodrigues2011structural}. The SST framework supports modeling of the components and interfaces for conventional computing and enables user-defined components, such as those for the QPU element introduced here. In addition, from these component models, SST supports discrete-event simulations to characterize and quantify the interactions between components by collecting statistics.
\par
We have used SST to develop a series of component models for the QCU and QEU that implement the QPU design. We do not implement the register components or the field language as our studies do not currently investigate verification of quantum register state, a key aspect of debugging quantum software. Rather we examine the system-level interactions that parses and transmit messages between components in order to ensure logical correction and resource management. Future studies are prepared to integrate these system models with quantum circuit simulation methods as part of software debugging tools. 
\par
As an example of how we use system-level modeling and simulation, we have simulated the execution of a test program based on quantum search. Quantum search is a well-known example of how quantum computing may accelerate the task of finding a marked element within an unstructured database. As originally proposed by Grover, quantum search requires $\sqrt{N}$ operations to recover the element from a data base of $N$ elements, whereas brute force search requires $N/2$ elements on average \cite{grover1996fast}. This quadratic speedup in the search speed is an interesting use case in scientific computing, for example, in identifying correlations between disparate data sets.
\par
Our implementation decomposes the quantum search algorithm into a sequence of one and two-qubit gate operations. This ISA is widely used by the quantum algorithms community but it is not strictly necessary. In addition, we have addressed aspect of scheduling resources by assuming a serial sequence of gate operations. This ensures that no register element or QEU can be simultaneously addressed by the QCU. However, this is an overly conservative schedule as many operations in our instruction list can be executed concurrently. In addition, a serial schedule greatly increases the duration of the quantum program. Because the computational state of the register decays due to decoherence and dephasing, this choice of schedule also greatly limits the depth of a realistic program. We address this concern by also including error correction into our simulation of the algorithmic methods. Quantum error correction and fault-tolerant operations enable the effects of noise to be overcome, but at the cost of increased qubit and gate resource requirements. We have used two levels of Steane [[7,1,3]] error correction alongside Shor ancilla and syndrome measurements.
\par
\begin{table}
\renewcommand{\arraystretch}{1.5}
\begin{center}
\caption{Gate models for QPU opcodes.}
\label{tab:ene}
\begin{tabular}{ | l | r | r | r|}
\hline
Gate & Duration & Power & Energy \\ \hline
INIT & 300 $\mu$s & 0.1 pW & 5 aJ \\
Unitary & 40 ns & 0.1 pW & 4 zJ \\
READ & 100 $\mu$s & 0.1 pW & 5 aJ \\
\hline
\end{tabular}
\end{center}
\end{table}
\par
A key part of the quantum search algorithm is the implementation of the database to be searched. The typical discussion invokes an oracular subroutine that applies some hidden method for verifying when a marked state has been located. We do not use an oracle in our implementation, but rather we implement the database as a multi-control \textsc{not} gate, which has the effect of marking the computational register with the sought-after state. Of course, the implementation of the database query requires us to have a priori information about what is the marked element, but we use this implementation for test purposes only. In total, the number of bare logical gates in the program scales as $O(2^{n/2})$, with $n = \log_2 N$. A single layer of quantum error correction incurs an additional overhead of $O(n^7)$ gates.
\par
Our SST model accepts as text-based input the string of program instruction to the host CPU. The QCU component is initialized with a microcode architecture that defines opcodes for each of the available instructions on the device. We assume the program input has already been decomposed into this ISA, so that the sequence of instructions set to the QCU from the CPU do not require further decomposition. The QCU parses these instructions and register addresses into the corresponding opcode and then dispatches it to the correct QEU. Note that our model for the QCU is relatively simple and it does not include the logical for routing and signaling opcodes to a QEU - this will be investigated later.
\par
The model for each QEU is to consume the received instruction by activating a sequence of field generators that applies the needed interactions onto the register. We do not simulate the explicit action of generating the fields, but rather we tally the amount of time required for each received gate in the program. In addition, we also tally energy consumption for each gate. The latter metrics depend on the specific technology and device design underlying the QPU. For the present study, we consider a silicon-based QPU derived from a recent design \cite{Tosi2017}. We summarize the results of this model in Table~\ref{tab:ene} but details of these calculations are available upon request. For this technology, the dominate energy consumption comes from the initialization and readout instructions, which arise during the program for the implementation of quantum error correction as well as for register preparation and measurements.
\par
\begin{figure}
\includegraphics[width=\columnwidth]{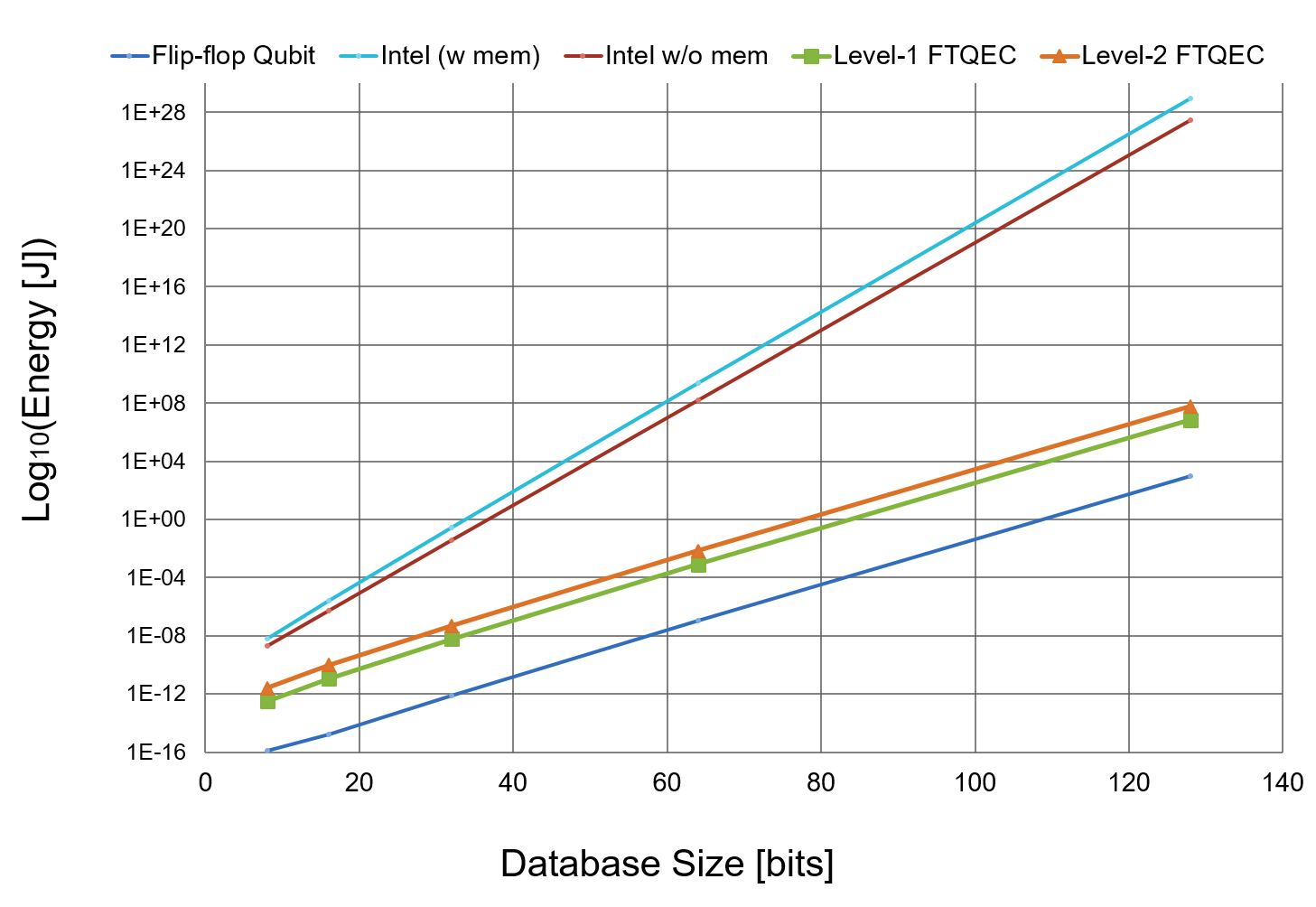}
\caption{The estimated minimal energy consumption for searching an unstructured database with respect to the size of the database. Lines represent conventional methods with and without memory costs as well as quantum search methods using a flip-flop silicon-based processor. These estimates only account for minimal amount of energy to perform the search operation. }
\label{fig:res}
\end{figure}

For the quantum search program, we have tallied the total energy needed to perform a search over $N$ items. The results of these simulations are shown in Fig.~\ref{fig:res}. We have compared these calculations to similar models for brute force search using a conventional processor (Intel i7). The conventional models account for only the $n$-bit comparator required to identify the marked item within the database and the main memory transfers required to fetch $N/2$ elements. It is apparent that all of the models scale exponentially with the size of the data base, as expected based on their algorithmic complexities. However, our models for the quantum search program suggest energy requirements that are order of magnitudes lower than conventional processing. This is true even in the presence of multiple layers of quantum error correction. 
\par
We must emphasize that our models for QPU and CPU program execution have excluded many significant sources of energy consumption. Our purpose for these simulations has been to emphasize the minimal energy consumption of lowest-level instructions and opcodes issued within the device model. For quantum computing, it is understood that manipulating individual electrons, as is the case for silicon-based technologies, requires very little energy. These physical limitations are not met in practice, however, because current technology for the field generators used to actuate these gates are typically based on macroscopic fields. We are refining our model to include these additional sources of energy as well. However, our model for the CPU has also emphasized the minimum energy per instructions required. We have not accounted for cooling and other inefficiencies in either model.
\section{Conclusions}
We have presented an overview of the structure and operation of quantum accelerators for high-performance computing. Our analysis has focused on only a single QPU-accelerated node, which may be a reasonable assumption in the near term. We have demonstrated how modeling and simulation at the system level provides insights into both the structure and behavior of these accelerated nodes. However, additional questions arise regarding how quantum acceleration overlaps with traditional parallelism and we expect to explore this question with future models. 
\section*{Acknowledgments}
This manuscript has been authored by UT-Battelle, LLC, under Contract No. DE-AC0500OR22725 with the U.S. Department of Energy. The United States Government retains and the publisher, by accepting the article for publication, acknowledges that the United States Government retains a non-exclusive, paid-up, irrevocable, world-wide license to publish or reproduce the published form of this manuscript, or allow others to do so, for the United States Government purposes. The Department of Energy will provide public access to these results of federally sponsored research in accordance with the DOE Public Access Plan (http://energy.gov/downloads/doe-public-access-plan).

\bibliographystyle{IEEEtran}
\bibliography{icrc2017}

\end{document}